\documentclass[12pt]{article}
\usepackage{fullpage,graphicx,psfrag,amsmath,amsfonts,verbatim}
\usepackage[small,bf]{caption}\usepackage{amssymb}
\usepackage{enumitem}
\usepackage{tikz}
\usepackage{listings}
\usepackage{authblk}
\usepackage{textcomp}
\graphicspath{{figures/}}

\definecolor{codegreen}{rgb}{0,0.6,0}
\definecolor{codegray}{rgb}{0.5,0.5,0.5}
\definecolor{codepurple}{rgb}{0.58,0,0.82}
\definecolor{backcolour}{rgb}{0.95,0.95,0.98}

\lstdefinestyle{mystyle}{
    backgroundcolor=\color{backcolour},   
    keywordstyle=\color{magenta},
    numberstyle=\tiny\color{codegray},
    stringstyle=\color{codepurple},
    basicstyle=\ttfamily\small,
    breakatwhitespace=false,         
    breaklines=true,                 
    captionpos=t,                    
    keepspaces=true,                 
    numbers=left,                    
    numbersep=5pt,                  
    showspaces=false,                
    showstringspaces=false,
    showtabs=false,                  
    tabsize=2,
}

\definecolor{seagreen}{rgb}{0.18, 0.55, 0.34}
\definecolor{mediumviolet-red}{rgb}{0.78, 0.08, 0.52}
\definecolor{khaki}{rgb}{0.94, 0.9, 0.55}

\lstdefinelanguage{mypython}
{
	keywords=[1]{from, import, as, assert, not, print, nonneg, PSD},
	keywordstyle=[1]{\color{mediumviolet-red}},
	keywords=[2]{cvxpy,Variable, 
	sqrt, exp, numpy, np, Problem, Minimize, Maximize, value, solve,
    sum, multiply, arange, range, norm1, norm2, norm_inf, abs, square,
    diagonal, outer},
	keywordstyle=[2]{\color{seagreen}},
	upquote=true,
	showstringspaces=false,
	basicstyle=\ttfamily,
	columns=fullflexible,
	keepspaces=true,
	emph={True,False,def,return,float,class,match,switch,len},
	emphstyle={\color{seagreen}},
	belowskip=1em,
	aboveskip=1em,
    morecomment=[l]{\#}
}

\usepackage{hyperref}
\usepackage{wasysym}
\hypersetup{
colorlinks   = true,    
urlcolor     = blue,    
linkcolor    = blue,    
citecolor    = red      
}

\usepackage[
backend=bibtex,
style=alphabetic,
sorting=ynt,
firstinits=true,
url=false, doi=false, eprint=false,
]{biblatex}
\usepackage{color}
\newcommand{\reals}{{\mbox{\bf R}}}

\newcommand{\Tr}{\mathop{\bf Tr}}

\newcommand{\Prob}{\mathop{\bf Prob}}

\newcommand{\argmin}{\mathop{\rm argmin}}

\newcommand{\sign}{\mathop{\bf sign}}

\newcommand{\eg}{{\it e.g.}}
\newcommand{\ie}{{\it i.e.}}

\newcommand{\BEAS}{\begin{eqnarray*}}
\newcommand{\EEAS}{\end{eqnarray*}}
\newcommand{\BEA}{\begin{eqnarray}}
\newcommand{\EEA}{\end{eqnarray}}
\newcommand{\BEQ}{\begin{equation}}
\newcommand{\EEQ}{\end{equation}}
\newcommand{\BIT}{\begin{itemize}}
\newcommand{\EIT}{\end{itemize}}

\newcounter{algorithmctr}
\renewcommand{\thealgorithmctr}{\arabic{algorithmctr}}
   {\mbox{}\\*[\parskip]\begin{minipage}{\linewidth}%
       \refstepcounter{algorithmctr}\begin{list}{}{%
       \setlength{\rightmargin}{0\linewidth}%
       \setlength{\leftmargin}{.05\linewidth}}%
       \rmfamily\small
       \item[]{\setlength{\parskip}{0ex}\hrulefill\par%
        \nopagebreak{\bfseries\textsf{Algorithm \thealgorithmctr~}}}}%
   {{\setlength{\parskip}{-1ex}\nopagebreak\par\hrulefill\\*[2ex]\par}%
   \end{list}\end{minipage}}

\title{Exponentially Weighted Moving Models}

\author[1]{Eric Luxenberg}
\author[1]{Stephen Boyd}
\affil[1]{Department of Electrical Engineering, Stanford University}

\begin{document}
\maketitle

\begin{abstract}
An exponentially weighted moving model (EWMM) for a vector time series fits a 
new data model each time period, based on an exponentially fading loss function
on past observed data.  The well known and widely used
exponentially weighted moving average (EWMA) is a special case that estimates
the mean using a square loss function.
For quadratic loss functions EWMMs can be fit using a simple recursion 
that updates the parameters of a quadratic function.  For other 
loss functions, the entire past history must be stored, and the fitting problem
grows in size as time increases.
We propose a general method for computing an approximation of 
EWMM, which requires storing only a window of a fixed number of past samples, and uses
an additional quadratic term to approximate the loss associated with 
the data before the window.
This approximate EWMM relies on convex optimization, and solves problems that do
not grow with time.  We compare the estimates produced by our approximation
with the estimates from the exact EWMM method.
\end{abstract}

\newpage
\tableofcontents
\newpage

\section{Introduction}

We consider the problem of fitting a time-varying model to a vector time series,
updating it each time period as new data is observed.
Assuming that recent data is more relevant than data from many periods
in the past, the model is fit giving more weight to recent past values
and lower weight to values far in the past.

\paragraph{Rolling window model.}
One simple method to do this is to fit the model at time period $t$
using a rolling window of $R$ previous values of the time series.
The choice of $R$ involves a trade-off.  When it is small, we have fewer data
to fit our model; when it is large, the model takes longer to adapt to 
changes in the underlying data.
We can think of a rolling window model (RWM) with window length $R$ as one that puts weight
one on the last $R$ data values, and weight zero on any values more than $R$ periods in
the past.  
One advantage of such an RWM is that the optimization problem we solve 
to carry out the fitting has the same size in each time period.

\paragraph{Exponentially weighted moving model.}
Another method for fitting a time-varying model
uses all past data to create the model, but puts a time-varying weight on past
values that decays smoothly as we move farther back in time.  A natural
choice for the weights is an exponential decay.  We refer to such a model
as an exponentially weighted moving model (EWMM).
The parameter in EWMM analogous to $R$ in an RWM is the half-life, the 
number of periods in the past where the weight decays to one-half.

\paragraph{Exponentially weight moving average.}
EWMMs generalize the well known and widely used exponentially weighted moving 
average (EWMA).  When we fit the data with a constant model using a square 
loss function, \ie, we attempt to estimate the mean, EWMM reduces to EWMA.
But EWMM includes many other interesting data models beyond EWMA,
such as exponentially weighted quantile estimation, exponentially weighted covariance
estimation, and various exponentially weighted regression models, possibly with
regularization.

\paragraph{Fixed size recursion.}
One attractive property of the EWMA estimate is that while it is based on
all past values, it can be computed recursively, so there is no need to store
all past data, and the computational effort to compute the EWMA estimate 
does not grow with time.  This is similar to an RWM, where a fixed size problem
is solved in each time period.

For EWMMs with quadratic loss functions a similar recursion can be used to
fit the model.  A well known example is exponentially weighted least squares.
We will see that other more complex models can be handled using this method.
For example, we can fit an exponentially weighted sparse inverse covariance matrix
to a vector time series, with a fixed amount of storage and computation each 
step.

\paragraph{Approximate recursive method.}
For a general EWMM, there is no simple recursion that allows us to exactly
evaluate the EWMM estimate; we must store all past values and then solve
a problem that grows with time.
This leads us to our focus in this paper: Approximate recursive methods,
which store only a fixed size window of past data, and solve a problem 
of constant size in each period. Such methods compute approximations to
the true EWMM estimates.
The key to these approximations is to form a tractable approximation
of the loss function corresponding to data that falls in the tail,
outside the fixed size window we keep.  We do this using a quadratic 
tail approximation.

\paragraph{This paper.}
We introduce the concept of a general EWMM, focussing on models
that can be fit via convex optimization.  
We address the question of how to compute an EWMM.  For quadratic loss functions,
there is an exact recursive method.
For nonquadratic losses, we describe methods to approximately fit 
an EWMM, using only a fixed 
window of past values, and solving a problem of fixed size that does not 
grow with time.  
We demonstrate our approximate recursive method with examples,
showing that it finds models that are close to the exact EWMM models,
fit using all past data.
The method is practical, and extends the many 
data models that are fit using convex optimization to the exponentially 
time-varying setting.

In this paper we do not address the question of whether an EWMM should be used in 
an application, for example
instead of an RWM or any other method for fitting a time-varying model.
We simply assume that a user wishes to use an EWMM, and give a practical method
to carry out the evaluation with memory and computation that do not grow 
with time.

\subsection{Previous and related work}
\paragraph{EWMA.}
The idea of the exponentially weighted moving average (EWMA) is well known, and
has origins going as far back as the recursive exponential window functions 
used by Poisson in the 19th century. 
It was introduced to statistics in 1956 by Brown as
a method for forecasting demand \cite{adl1956exponential}. In the context of
signal processing, the EWMA is an application of a window function, used as a
low pass filter to remove high frequency noise from a signal
\cite{oppenheim1999discrete}.
Moving averages are related to the concept of rolling window estimation, wherein
models are repeatedly fit to a fixed size window of past data. RWMs are
widely used in time series analysis and forecasting in economics, finance, and
engineering \cite{box2015time,tsay2005analysis}. 

Exponentially weighted sums also appear in finance, albeit applied to future
rather than past terms. Cash flows are discounted in the calculation of present
value; here the discount factor is $\frac{1}{1+r}$, where $r$ is the interest
rate. Exponential weighting of future terms also appears in Markov decision
processes (MDPs) \cite{bertsekas2022abstract}, where the value function represents the discounted sum of
future rewards. Exponentially weighted moving averages appear in the context of
model free reinforcement learning, where an exponentially decayed expectation of
rewards is incrementally estimated via a recursive update
\cite[\S17.1-3]{kochenderfer2022algorithms}.

\paragraph{Online quantile estimation.}
In online quantile estimation,
the goal is to estimate the $\eta$-quantile of a time series, for some
fixed $\eta \in (0,1)$.  This can be computed exactly at any time by
storing all past data and sorting it, but this need not be practical for 
large data sets.  Several methods
exist for online quantile estimation, such as the $P^2$ algorithm for online
quantile estimation without storing all points \cite{jain1985p2}. This method
stores only 5 points chosen on the empirical CDF. Another well known work by
Greenwald and Khanna \cite{greenwald2001space} provides a space efficient method 
based on defining a notion of approximate quantiles, where the approximation
error is allowed to grow with the number of data points.

\paragraph{Moving regression.}
The EWMM is a generalization of the exponentially weighted moving regression
model, which has been studied in the context of time series forecasting. The
original work on this topic is due to \cite{christiaanse1971short}, who extended
the method described by Brown \cite{adl1956exponential} to the case of 
linear regression, which allowed
for the use of features in the forecasting model.
Another related area of work is locally weighted regression, which also fits a
regression model with a weighted sum of error functions
\cite{cleveland1979robust}. The weights are typically chosen to be a function of the 
distance between the point at which a prediction is being made and the data
points being used to fit the model. In our setting, the notion of distance is
temporal rather than spatial, and the weight decreases exponentially
with distance, \ie, lapsed time.
Hastie and Tibshirani \cite{tibshirani1987local} provide a treatment of local likelihood
estimation, which generalizes moving window linear regression to 
likelihood based regression models, where maximum likelihood estimates are 
computed on a window of data points near the point at which a prediction is 
being made.
In the case of moving window linear regression, a classic algebraic trick that
allows for reduced computational complexity of the recursive update of estimates
has been known since Gauss and Legendre
\cite{sorenson1970least,gauss1995theory}.

\paragraph{Online learning.}
In computer science, online learning concerns the problem of
learning from a stream of data, where the goal is to make predictions about the 
next data point, and update the model based on the observed data. This is in
contrast to batch learning, where the model is trained on a fixed set of data. 
See \cite{mcmahan2017survey} for a survey of online learning algorithms. Hazan
provides a comprehensive overview of online convex optimization in
\cite{hazan2016introduction}. 
The exponentially weighted moving model can be considered a form of online
learning. However, much of the work in online learning focuses on the
regret, the difference between the performance of the 
online learning algorithm and the best fixed model in hindsight. The motivation
for EWMM differs in that our goal is to estimate a time-varying parameter well
at each time period. 
But like EWMM, online learning solves a problem, typically convex and of fixed
size, each time period to update its estimates of the parameters.

\paragraph{Quadratic surrogates and tail approximations.}
The idea of approximating part of an objective function as a convex
quadratic is a basic one in several optimization methods,
most famously Newton's method \cite[\S9.5]{cvxbook}.
A sophisticated extension is
sequential quadratic programming (SQP) methods \cite{boggs1995sequential}. 
In the context of control, quadratic approximations of the value function lead
to approximate dynamic programming (ADP methods) \cite{powell2007approximate,keshavarz2014quadratic}.
A convex quadratic terminal cost or value function in often used in model predictive control
(MPC) \cite{wang2009performance,wang2015approximate}.

\subsection{Outline}
In \S\ref{s-ewmm} we formally describe EWMMs, and 
in \S\ref{s-quad-ewmm} we consider the special case when the loss function
is quadratic.
In \S\ref{s-approx-ewmm} we describe methods for approximating 
an EWMM using quadratic tail approximations.
We give some numerical examples using both synthetic and real data
in \S\ref{s-examples}.

\section{Exponentially weighted moving model}\label{s-ewmm}
\subsection{Exponentially weighted moving average}
Suppose $x_1, x_2, \ldots \in \reals^n$ is a vector time series.
Its \emph{exponentially weighted average} (EWMA) is the vector time series
\BEQ\label{e-ewma}
\tilde x_t = \alpha_{t} \sum_{\tau=1}^{t}\beta^{t-\tau} x_\tau,
\quad t=1,2, \ldots,
\EEQ
where $\beta \in(0,1)$ is the forgetting factor, and
\BEQ\label{e-normalization}
\alpha_t=\left( \sum_{\tau=1}^{t}\beta^{t-\tau} \right)^{-1} 
= \frac{1-\beta}{1-\beta^{t}}
\EEQ
is the normalization constant.
The forgetting factor $\beta$ is usually
expressed in terms of the half-life $H=-\log 2 / \log\beta$, for which 
$\beta^H =1/2$. 

\paragraph{Recursive implementation.}
The EWMA sequence \eqref{e-ewma} can be computed recursively as
\BEQ\label{e-ewma-recursive} 
 \tilde x_{t+1}= \frac{\alpha_{t+1}}{\alpha_t} \beta
 \tilde x_t + \alpha_{t+1} x_{t+1}, \quad t=1,2, \ldots.
\EEQ
Thus we can compute $\tilde x_t$ without storing the past values
$x_1, \ldots, x_t$; we only need to keep track of the state $\tilde x_t$.

\paragraph{Interpretations.}
There are several ways to interpret the EWMA time series $\tilde x$.  We can think
of it as a version of the original time series $x$ which has been smoothed 
over a timescale on the order of $H$.
We can think of the transformation from the sequence $x$ to the EWMA 
sequence $\tilde x$ as a 
low-pass filtering operation, which removes high frequency variations.

The interpretation most useful in this paper is that $\tilde x_t$ is a 
time-varying estimate of the mean of $x_t$, formed from $x_1, \ldots, x_t$,
where we imagine that $x_t$ comes from a time-varying distribution with slowly
varying mean.
We can express this interpretation using a quadratic loss function:
\BEQ\label{e-ewma-loss}
\tilde x_t = \argmin_x \alpha_t \sum_{\tau=1}^t \beta^{t-\tau} \| x-x_\tau\|_2^2.
\EEQ
So the EWMA estimates $\tilde x_t$ minimize the exponentially weighted sum 
of previous quadratic losses $\|x-x_\tau\|_2^2$, $\tau=1,\ldots, t$.

\subsection{Exponentially weighted moving model}
The EWMM is a generalization of EWMA,
specifically the exponentially weighted loss formulation \eqref{e-ewma-loss}.
We consider a model of the data $x\in \reals^n$ that is parametrized 
by $\theta \in \Theta \subseteq \reals^m$, and specified by the loss function
$\ell: \reals^n \times \Theta \to \reals$, which we assume is convex in $\theta$.
(In particular, we assume $\Theta$ is a convex set.)
We interpret $\ell(x;\theta)$ as a measure of mis-fit with the data value $x$
and parameter value $\theta$, with small values meaning
the data $x$ is consistent with the model with parameter $\theta$.
The exponentially weighted loss at time $t$ is given by
\[
\alpha_t \sum_{\tau=1}^t \beta^{t-\tau}\ell(x_\tau;\theta),
\]
where $\beta\in (0,1)$ is the forgetting factor and $\alpha_t$ is the normalization
constant \eqref{e-normalization}.
The time-varying EWMM estimate of the parameter is given as
\BEQ \label{e-ewmm}
\theta_t = \argmin_{\theta\in \Theta} \left(
    \alpha_t \sum_{\tau=1}^t \beta^{t-\tau}\ell(x_\tau;\theta) + r(\theta) \right),
\EEQ
where $r: \Theta \to \reals \cup\{\infty\}$ is a convex regularizer.
This is a convex optimization problem, and so, computationally tractable.
We will assume that there is at least one minimizer in
the argmin above; if there are multiple minimizers, we can simply choose one.
We can see that with quadratic loss $\ell(x;\theta)=\|\theta-x\|_2^2$ and 
zero regularizer $r=0$,
EWMM reduces to EWMA (with $\theta_t=\tilde x_t$).

With one general exception described below, 
the EWMM cannot be computed recursively,
as in EWMA; to compute $\theta_t$ we generally need to 
store the entire set of past data $x_1, \ldots, x_t$.
Moreover the convex optimization problem we must solve to evaluate
$\theta_t$ grows in size with $t$.
Under the most favorable circumstances the complexity of solving the
problem involving all past data grows linearly with $t$; it follows 
that the computational complexity of computing $\theta_1, \ldots, 
\theta_t$ grows at least quadratically in $t$.


\subsection{Examples} Here we list some well known examples.
We start with data models that assume $x_t$ are 
independent samples from a fixed distribution family, with slowly
varying parameter $\theta_t$. 
We first describe examples with scalar $x_t$, for simplicity.

\subsubsection{Exponentially weighted moving data models}

\paragraph{Robust mean estimator.}
Instead of quadratic loss we can use a robust loss function such as
the Huber loss, which would give the exponentially weighted moving 
robust estimate of the mean \cite{huber1992robust}.

\paragraph{Quantile estimator.}
With loss $\ell(x;\theta) =|\theta-x|$, we obtain 
the exponentially weighted moving estimate of the median.
More generally using pinball or quantile loss 
\BEQ\label{e-pinball-loss}
\ell(x;\theta) = \max\{(1-\eta)(\theta-x), \eta (x-\theta)\},
\EEQ
where $\eta \in [0,1]$ is the quantile level,
we obtain the exponentially weighted moving estimate of the $\eta$-quantile
\cite{koenker1978regression}. 

\subsubsection{Exponentially weighted moving regression models}
The examples above fit moving models to the data $x_t$.
The same general form can also be used to fit regression or prediction
models as well.
Here we partition $x_t$ into two vectors, $x_t=(z_t,y_t)$ and 
seek a regression model, parametrized by $\theta$,
that predicts $y_t$ given $z_t$, parametrized as $\hat y_t = \theta z_t$.
(Here $z_t$ is the feature vector, and $y_t$ is the target.)
As a special case, suppose that $z_t=(y_{t-1}, \ldots, y_{t-M})$,
\ie, the feature vector consists of the previous $M$ values 
of $y_t$.  This gives us an exponentially weighted auto-regressive (AR)
prediction model.

\paragraph{Regression.}
We use loss $\ell(x,\theta) = L(y_t - \hat y_t)$, where $L$ is a 
convex loss function.  With $L(u)=\|u\|_2^2$, we get the exponentially 
weighted ordinary least squares regression model.
We can use other losses such as pinball or Huber.
We can add any convex regularization.
With regularizer $r(\theta)= \lambda \|\theta\|_2^2$, where $\lambda>0$ is 
a hyper-parameter, we obtain exponentially weighted ridge regression
\cite[page~564]{golub2013matrix}
With $r(\theta) = \lambda \|\theta\|_1$, we obtain the exponentially 
weighted LASSO regression model \cite{tibshirani1996regression}.  
With $r(\theta)=0$ for $\theta\geq 0$
(elementwise) and $r(\theta)=\infty$ otherwise, we obtain
the exponentially weighted nonnegative least squares regression model.

\paragraph{Logistic regression.}
With Boolean target data, \ie, $y_t \in \{-1,1\}$, and loss 
function 
\BEQ\label{e-logistic-loss}
\ell((z,y); \theta) = \log (1+\exp -yz^T\theta)
\EEQ
we obtain
exponentially weighted logistic regression \cite{hastie2009elements}.

\section{EWMM with quadratic loss}\label{s-quad-ewmm}
When the loss function $\ell$ is quadratic (including a linear 
and constant term), we can compute the EWMM parameter using a simple recursion 
similar to EWMA, storing only a fixed-size state and carrying out 
computations of constant complexity.

A general quadratic loss has the form
\[
\ell(x;\theta) = (1/2) \theta ^T P(x)\theta + p(x)^T\theta  + \pi(x),
\]
where $P(x)$ is positive semidefinite.
The exponentially weighted loss is also a convex quadratic function,
\[
\alpha_t \sum_{\tau=1}^t \beta^{t-\tau} \ell(x_\tau;\theta) =
(1/2) \theta ^T P_t\theta + p_t^T\theta  + \pi_t, 
\]
with
\[
P_t = \alpha_t \sum_{\tau=1}^t \beta^{t-\tau} P(x_\tau), 
\quad p_t = \alpha_t \sum_{\tau=1}^t \beta^{t-\tau} p(x_\tau),
\quad \pi_t = \alpha_t \sum_{\tau=1}^t \beta^{t-\tau} \pi(x_\tau).
\]

\subsection{Recursion for quadratic loss}
A simple recursion allows us to store $P_t$, $p_t$, and $\pi_t$
and update them as new data arrives, via 
\BEAS
P_{t+1} &=& \frac{\alpha_{t+1}}{\alpha_t}\beta P_t + \alpha_{t+1} 
P(x_{t+1}),\\
p_{t+1} &=& \frac{\alpha_{t+1}}{\alpha_t}\beta p_t + 
\alpha_{t+1} p(x_{t+1}),\\
\pi_{t+1} &=& \frac{\alpha_{t+1}}{\alpha_t}\beta p_t + 
\alpha_{t+1} \pi(x_{t+1}).
\EEAS
To find the EWMM parameter we solve the fixed-size convex optimization 
problem of minimizing
\[
(1/2) \theta ^T P_t\theta + p_t^T\theta  + \pi_t +r(\theta).
\]
(Since $\pi_t$ is a constant, it can be dropped.)

\subsection{Examples}\label{s-quad-ewmm-examples}

\paragraph{Ridge regression.}
We can also add a convex regularizer to the loss function, such as
$r(\theta) = \lambda \|\theta\|_2^2$, where $\lambda>0$ is a hyper-parameter.
This gives the exponentially weighted moving ridge regression model.

\paragraph{Lasso.}
We can easily extend to other penalties, such as the lasso penalty 
$r(\theta) = \lambda \|\theta\|_1$,
where $\lambda>0$ is a hyper-parameter. 

\paragraph{Nonnegative least squares.}
We use regularizer $r(\theta)=0$ for $\theta \geq 0$ (elementwise)
and $r(\theta) = \infty$ otherwise.

\paragraph{Gaussian covariance estimator.}
We model vector data as $x_t \sim \mathcal N(0,\Sigma_t)$.
We parametrize the model using
$\theta_t =\Sigma_t^{-1}$, the symmetric positive definite precision matrix.
To form the exponentially weighted covariance estimate, 
we minimize the convex function
\[
\alpha_t \sum_{\tau=1}^t \beta^{t-\tau} \left(x_\tau^T\theta x_\tau - 
\log \det \theta\right),
\]
which is the weighted negative log likelihood, with a factor of one-half
and an additive constant.
We express this as
\[
\Tr \left(\alpha_t \sum_{\tau=1}^t \beta^{t-\tau} x_\tau x_\tau^T\right)\theta
- \log \det \theta.
\]
The first term is linear in $\theta$, and therefore also quadratic.
We take this linear term as our loss and
\BEQ\label{e-log-det-theta}
r(\theta) = -\log \det \theta 
\EEQ
as our regularizer,
even though the log determinant term is also typically considered part of the loss.
With this re-arrangement the EWMM has quadratic loss, so we can use 
the recursion above to solve it exactly by solving a fixed-size convex problem.
It is not hard to show that the EWMM estimate is
the traditional exponentially weighted empirical covariance estimate,
\[
\theta_t^{-1} = \alpha_t \sum_{\tau=1}^t \beta^{t-\tau} x_\tau x_\tau^T
\]
(see, \eg, \cite{menchero2011barra,johansson2023covariance}).

\paragraph{Sparse inverse covariance estimator.}
To obtain a sparse precision matrix,
we add $\ell_1$ regularization on the off-diagonal entries to the
regularizer \eqref{e-log-det-theta},
\[
\lambda \sum_{i \neq j} |\theta_{ij}|,
\]
with $\lambda >0$ \cite{friedman2007sparse}.
We recursively compute the EWMA empirical estimate 
\[
\Sigma_t^\text{emp} =
\alpha_t \sum_{\tau=1}^t \beta^{t-\tau} x_\tau x_\tau^T,
\]
and then obtain the EWMM estimate $\theta_t$ as the minimizer of
\[
\Tr \left(\Sigma_t^\text{emp}\right)^T \theta - \log \det \theta 
+ \lambda \sum_{i \neq j} |\theta_{ij}|.
\]

\paragraph{Probability mass estimator.}
Suppose that $x_t$ takes on only the values $1, \ldots, m$,
and we wish to estimate the
probability mass function (PMF) parametrized as 
\[
\Prob(x=k) = p_k = \frac{\exp \theta_k}{\exp \theta_1 + \cdots + \exp \theta_m},
\quad k=1, \ldots, m,
\]
with $\theta \in \reals^m$.
(To remove the redundancy in the parameterization we can add the convex constraint
$\theta_1 + \cdots + \theta_m=0$.)
We use negative log-likelihood loss,
\[
\ell(x,\theta) = - \theta_k + \log\left(\exp \theta_1+ 
\cdots + \exp\theta_m\right), \quad x=k.
\]
(The subscripts on $\theta$ here denote entries, not time period.)

As simple regularizer is $r(\theta)=\lambda \| \theta \|_2^2$
where $\lambda>0$ is a hyper-parameter.
If the values $1, \ldots, m$ are nodes of a graph with weights $W_{ij}$ on the
edge between nodes $i$ and $j$, we can add Laplacian
regularization 
\[
r(\theta) = \frac{\lambda}{2}\sum_{i,j=1}^m W_{ij}(\theta_i-\theta_j)^2
\]
to obtain an exponentially weighted PMF estimate that is smooth with respect
to the graph \cite{tuck2021fitting}.

To get the exponentially weighted PMF estimate we minimize the convex function
\[
\alpha_t \sum_{\tau=1}^t \beta^{t-\tau} \ell(x_\tau; \theta) + r(\theta) = 
- \left( \alpha_t \sum_{\tau=1}^t \beta^{t-\tau} \left(e_{x_\tau}\right) \right)^T 
\theta
+ \log \left( \exp \theta_1+ \cdots + \exp \theta_m \right) + r(\theta),
\]
where $e_j$ is the standard $j$th unit vector in $\reals^m$, \ie,
$(e_j)_k=1$ if $k=j$ and $(e_j)_k=0$ if $k\neq j$.
The first term on the righthand side is linear in $\theta$, 
and therefore also quadratic, do we take that as our loss.
We take the second and third terms on the righthand side as the regularizer.
The vector in parentheses in the first term on the righthand side 
is the EWMA estimate of the past frequencies of
occurrence, which of course can be computed recursively.

Without regularization, it is easily shown that
the EWMM estimate is
\[
p_t=
\alpha_t \sum_{\tau=1}^t \beta^{t-\tau} \left(e_{x_\tau}\right),
\]
the EWMA frequencies of occurrence. (This assumes that each value has occurred
at least once.)


\paragraph{Exponential family.}  Some of the examples above are special cases of 
parameter estimation in an exponential family.
An exponential family of densities on $\reals^n$, with parameter 
$\theta\in \reals^m$, has the form
\[
p(x;\theta) = h(x)\exp(\theta^T T(x) - A(\theta)),
\]
where $T:\reals^n \to \reals^m$ is the sufficient statistic, $A:\reals^m \to
\reals$ normalizes the density, and $h:\reals^n \to \reals$ is the base measure.
It is well known that $A$ is convex.
Using the negative log-likelihood loss
\[
\ell(x;\theta) = -\log h(x) -\theta^T T(x) + A(\theta),
\]
the EWMM estimate of the parameter $\theta_t$ is the minimizer of
\[
\alpha_t \sum_{\tau=1}^t \beta^{t-\tau} \ell(x_\tau; \theta) +r(\theta)=
-\left(\alpha_t \sum_{\tau=1}^t \beta^{t-\tau} T(x_\tau)\right)^T\theta + A(\theta)
+r(\theta).
\]
(We drop $-\log h(x)$ since it does not depend on $\theta$.)
The first term on the righthand side is linear in $\theta$, and so 
can be computed recursively.  We only need to keep track of
the exponential weighted average of the sufficient statistic,
$\alpha_t \sum_{\tau=1}^t \beta^{t-\tau} T(x_\tau)$. 

The fact that the EWMM for exponential families can be computed with a finite
size problem is connected to a well known result in statistics, the 
Pitman-Koopman-Darmois theorem \cite{pitman1936sufficient,
koopman1936distributions, darmois1935lois}. The theorem says that, under some
minor technical conditions, the exponential family of distributions is the only
family where there can be a sufficient statistic whose dimension does not grow
with the sample size.

\section{Approximate finite memory EWMM}\label{s-approx-ewmm}

\subsection{Quadratic approximation of tail loss}\label{s-quad-approx}
We will explore methods that at time period $t$ store $x_{t-M}, \ldots, x_t$,
\ie, the current and previous $M$ data values.  (The parameter $M$ is called the memory.)
To motivate our approximate method, we first write the EWMM \eqref{e-ewmm} as
\BEQ\label{e-ewmm-tail}
\theta_t = \argmin_{\theta\in \Theta} \left(
    \alpha_t \sum_{\tau=t-M}^t \beta^{t-\tau}\ell(x_\tau;\theta) + V_t(\theta)+
r(\theta) \right),
\EEQ
where $V_t$ is the tail loss, defined as 
\BEQ\label{e-V}
V_t(\theta) = 
    \alpha_t \sum_{\tau=1}^{t-M-1} \beta^{t-\tau}\ell(x_\tau;\theta).
\EEQ
Note that in \eqref{e-ewmm-tail}, we only explicitly refer to the past $M+1$ 
data values $x_{t-M}, \ldots, x_t$, with the previous losses appearing implicitly in
the tail loss term $V_t(\theta)$.

Our approximation replaces $V_t(\theta)$ with a convex quadratic approximation
$\hat V_t(\theta)$, which gives the approximate EWMM 
\BEQ\label{e-ewmm-tail-approx}
\hat \theta_t = \argmin_{\theta\in \Theta} \left(
\alpha_t \sum_{\tau=t-M}^t \beta^{t-\tau}\ell(x_\tau;\theta) + \hat V_t(\theta)+
r(\theta) \right).
\EEQ

We will describe below two methods that can be used to form the tail approximation
$\hat V_t$ recursively, without storing the tail data $x_1, \ldots, x_{t-M-1}$.
Note that computing the approximate EWMM requires solving a problem of fixed size,
that does not grow with $t$.

\paragraph{Choice of $M$.}
The larger $M$ is, the closer our approximate EWMM parameter will be to the 
exact EWMM parameter, at the cost of solving a larger optimization problem.
When $M$ is larger than, say, $4H$, the tail contribution
is so small that the effect of $\hat V_t$ is very small, and any reasonable choice,
including $\hat V_t=0$, would likely give estimates very close to the exact EWMM
estimate.  So we are mostly interested in the case when $M$ is around $H$.

\subsection{Recursive Taylor approximation}\label{s-re-taylor}
Here we describe a method to construct the quadratic 
tail loss approximation $\hat V_t$ recursively
from $\hat V_{t-1}$ (which is quadratic) and $\ell(x_{t-M-1};\theta)$,
the loss term that joins the tail at time period $t$.
We start with the exact recursion, analogous to \eqref{e-ewma-recursive},
\[
V_{t}(\theta) = \frac{\alpha_{t}}{\alpha_{t-1}} \beta V_{t-1} (\theta) + \alpha_{t}
\ell(x_{t-M-1};\theta).
\]
We now replace $V_t(\theta)$ and $V_{t-1}(\theta)$ with their quadratic approximations
$\hat V_t(\theta)$ and $\hat V_{t-1}(\theta)$, and 
approximate the loss term $\ell(x_{t-M-1};\theta)$
with a convex quadratic approximation $\hat \ell(x_{t-M-1};\theta)$ to obtain
\BEQ\label{e-Vthat-update}
\hat V_{t}(\theta) = \frac{\alpha_{t}}{\alpha_{t-1}} \beta \hat V_{t-1} 
(\theta) + \alpha_{t} \hat \ell(x_{t-M-1};\theta).
\EEQ
This gives an explicit recursion for computing the quadratic tail loss 
approximation $\hat V_t$ from $V_{t-1}$ and $\ell(x_{t-M-1};\theta)$.
Note that we only need to store the coefficients of the quadratic functions.

It remains to specify the quadratic approximation of $\ell(x_{t-M-1};\theta)$.
We seek a convex quadratic approximation that is accurate near $\hat \theta_{t-1}$,
the previously computed parameter estimate.
When $\ell$ is twice differentiable with respect to $\theta$, 
an obvious approximation is its second-order 
Taylor expansion about the previous estimate,
\BEAS
\hat \ell(x_{t-M-1}; \theta) &=& 
\ell(x_{t-M-1};\hat \theta_{t-1})\\
&&~ + 
\nabla \ell(x_{t-M-1};\hat \theta_{t-1})^T (\theta-\hat \theta_{t-1})\\
&&~ + 
(1/2) (\theta-\hat \theta_{t-1})^T \nabla^2 \ell(x_{t-M-1};\theta_{t-1})
(\theta-\hat \theta_{t-1}),
\EEAS
where the gradient and Hessian are with respect to $\theta$.

When the loss has the form $\ell(x;\theta) = L(\theta^T x)$, where 
$L: \reals \to \reals$ is a convex loss function, the gradient and Hessian 
above have the simple forms
\BEAS
\nabla \ell(x_{t-M-1};\hat \theta_{t-1}) &=& L'(\hat \theta_{t-1}^T x_{t-M-1})
x_{t-M-1},\\
\nabla^2 \ell(x_{t-M-1};\hat \theta_{t-1}) &=& L''(\hat \theta_{t-1}^Tx_{t-M-1})
x_{t-M-1}x_{t-M-1}^T.
\EEAS

\subsection{Tail fitting}\label{s-re-evaluated-tail}

We now consider the case where $\ell$ is not twice differentiable, 
so we cannot use the
Taylor approximation to find the quadratic tail approximation $\hat V_t$.
In this case we can directly form a quadratic approximation of the tail.
To do this we store a second window of data within the tail,
\[
x_\tau, \quad \tau = t-M-M^\text{tail} , \ldots, t-M-1,
\]
where $M^\text{tail}$ is the additional memory we use to approximate
the tail, with $M+M^\text{tail}$ the total number of previous
values we must store.
The idea is to use these $M^\text{tail}$ points to form the quadratic
estimate $\hat V_t$, and the past $M$ values $x_{t-M}, \ldots, x_t$ to then 
form $\theta_t$.
This second window of past data is used purely for fitting the tail,
and so can potentially be much larger than $M$. This is because fitting the tail
approximation is typically cheaper than solving the EWMM problem of the same size.

\paragraph{Fitting the tail approximation.}
We propose the following procedure to fit the tail approximation $\hat
V_t(\theta)$ at time $t$.
\begin{enumerate}
\item Choose $L$ points $u_1,\ldots,u_L$ near $\theta_{t-1}$. 
\item Evaluate the tail losses at the $u_i$. For $i=1,\ldots,L$, let 
\[
v_i =
\alpha_t \sum_{\tau=t-M-M^\text{tail}}^{t-M-1} 
\beta^{t-\tau}\ell(x_\tau;u_i).
\]
\item Use least squares to fit a quadratic function parametrized by $P\in \reals^{m\times m}$,
$p\in \reals^m$, and $\pi\in \reals$ to the points $(u_i,v_i)$.
\end{enumerate}
To ensure that the quadratic approximation is convex, a constraint can be added
to the least squares problem to ensure that the $P$ is
positive semidefinite. This means the fitting problem is a semidefinite program
(SDP), which can increase the computational cost of fitting.
An alternative is to use least squares to find $\tilde P$,
and then simply project $P$ onto the set of
positive semidefinite matrices.  We have found this simpler method to be effective.

\paragraph{Generating evaluation points.} There are many principled methods to
generate the points $u_i$.
If the dimension of $\theta$ is small enough,
we can choose a uniform grid of points around $\theta_{t-1}$. 
One could also use a low discrepancy sequence
\cite{sobol1967distribution} as a more sophisticated way to cover the space.
One can also model the variance of $\theta$ across previous estimates and
generate points using so-called sigma points \cite{van2004sigma}.
Even simpler is to sample from a normal distribution centered at $\theta_{t-1}$.
See \cite[Chap.~13]{kochenderfer2019algorithms} for a detailed study of methods for
generating evaluation points for approximation problems.
Any method of generating the evaluation
points should exclude any points not in $\Theta$.

\paragraph{Default method.}
Although we have mentioned several potential methods for fitting the tail, we
suggest the following simple default method.
We suggest $M^\text{tail}\approx 3M$ as a good all-purpose choice.
A quadratic function of $\theta\in\reals^m$
has approximately $m^2/2+m$ parameters, so we suggest that $L$ should be 
a modest multiple of this number.
We recommend sampling from a normal distribution centered at $\theta_{t-1}$ as
it is simple and effective. We take the standard deviation as
$\sigma=\|\theta_{t-1}\|_2+\epsilon$, where $\epsilon$ is small.

\section{Numerical examples}\label{s-examples}
In this section we give some examples of evaluating the EWMM either exactly
(our first example) or approximately (for the others). All examples can
be reproduced using publicly available code at 
\url{https://github.com/cvxgrp/ewmm_code}.
We use CVXPY, a Python-embedded modeling language for convex optimization to
specify and compute the EWMM \cite{diamond2016cvxpy}.

\subsection{Sparse inverse covariance estimation}
We use the sparse inverse covariance model described in
\S\ref{s-quad-ewmm-examples} to estimate the covariance matrix of a time series of
daily financial returns. In this example we can compute the EWMM estimate
exactly using recursion. 

\paragraph{Data.}
We use the 10 Industry Portfolio dataset from Kenneth French's data library
\cite{frenchdata}. The dataset contains daily returns for 10 value-weighted industry
portfolios. We examine returns from the last 4 years, from 2020-01-02 to
2024-01-31 giving us data $x_1, \ldots, x_T \in \reals^{10}$, where $T=1027$. 

\paragraph{Parameters.}
We use a half-life of $H=63$ (one quarter). We evaluate the model for
$\lambda=2.5$, $5$, $7.5$, and $10$, which result in covariance 
estimates with inverses that are increasingly sparse.

\begin{figure}
    \centering
    \includegraphics[width=1.0\textwidth]{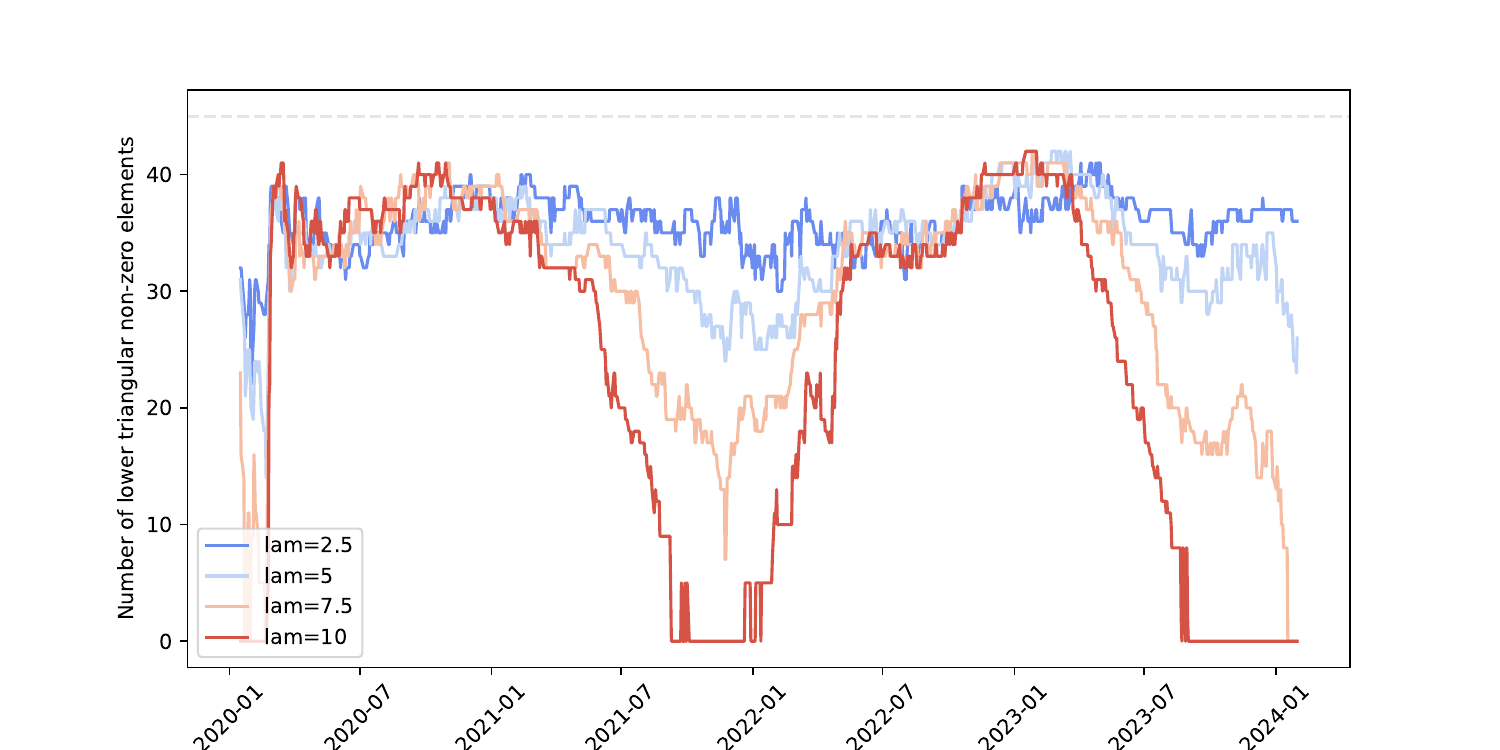}
    \caption{Number of nonzeros in the inverse covariance matrix across time for
    different values of $\lambda$.}
    \label{f-sic-sparsity}
\end{figure}

\begin{figure}
    \centering
    \includegraphics[width=1.0\textwidth]{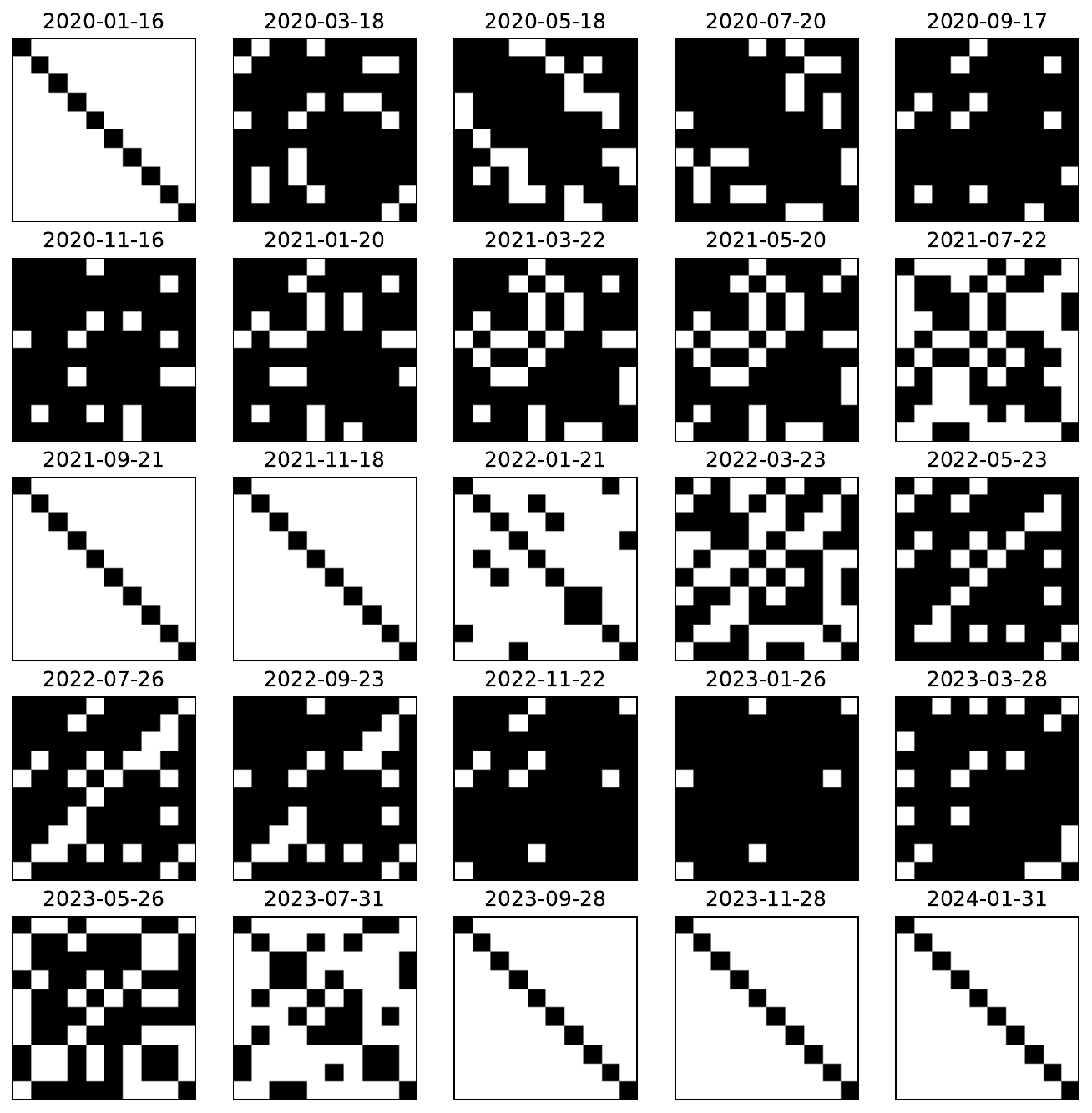}
    \caption{Sparsity patterns at various times for $\lambda=10$. White boxes
denote zero entries in the precision matrix.}
    \label{f-cov-estimates}
\end{figure}

\paragraph{Results.}
In figure~\ref{f-sic-sparsity} we show the sparsity of the inverse covariance
matrix across time for the different values of $\lambda$.  The plot 
gives the number of nonzero entries in the precision matrix,
with the dashed line at 45 showing the maximum possible value, \ie,
a fully dense precision matrix.
We also show examples of the inverse covariance matrix sparsity patterns 
at evenly spaced times in figure~\ref{f-cov-estimates}.
These plots show that the sparse inverse covariance estimate varies considerably 
with time, \ie, market conditions.

\begin{figure}
    \centering
    \includegraphics[width=1.0\textwidth]{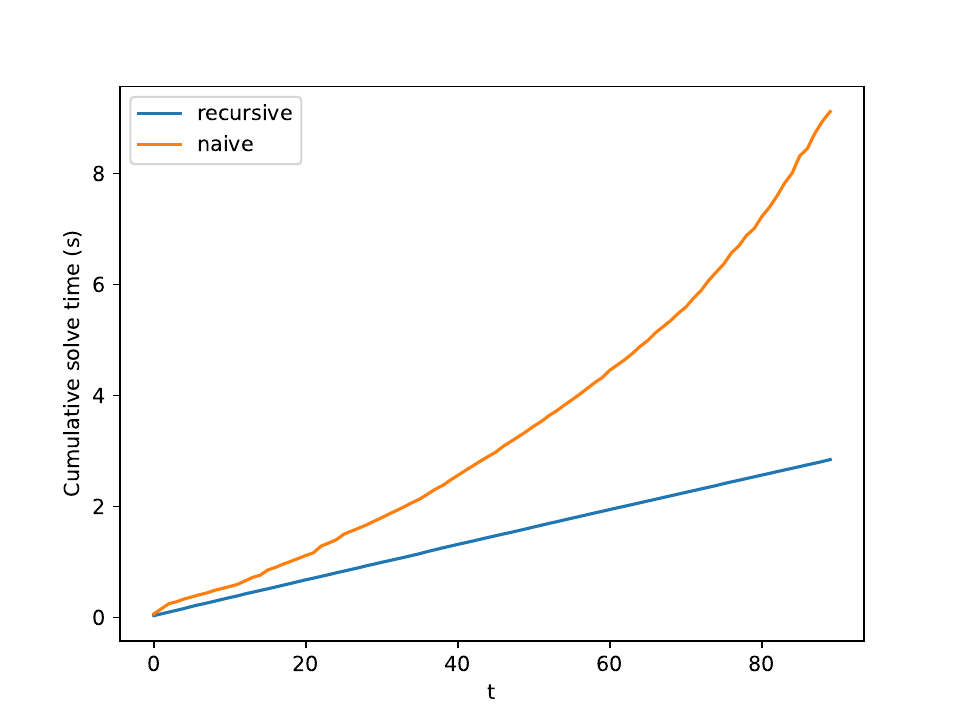}
    \caption{Cumulative time to fit the sparse inverse covariance estimation 
model using the na\"{i}ve method and the (exact) recursive method.}
    \label{f-cov-time}
\end{figure}

To illustrate the savings obtained from the recursive formulation,
we show the running computation time of fitting the model in 
figure~\ref{f-cov-time}.  As expected the na\"{i}ve method,
which saves all past data and directly computes the 
estimate using all past value, 
grows quadratically in time, whereas the recursive method grows linearly.

\clearpage 
\subsection{Quantile estimation}\label{s-quantile}
We use the pinball loss function \eqref{e-pinball-loss} to estimate 
the 15th, 50th, and 85th percentiles of a scalar time series. 
Since the pinball loss is not twice differentiable, we use the tail approximation
method described in \S\ref{s-re-evaluated-tail} to fit the tail using points
sampled from a normal distribution centered at the previous estimate with
standard deviation equal to one fifth of the magnitude of the previous estimate. 

\paragraph{Data.}
In this example we use synthetic data.
First we generate smoothly varying sequences
$\mu_t$ and $\sigma_t$ as
\BEAS
    \mu_t &=& a\sin(2\pi t/P_1) + b\cos(2\pi t/P_2),\quad t=1,\ldots,T,\\
    \sigma_t &=& c\sin(2\pi t/P_3) + d\cos(2\pi t/P_4), \quad t=1,\ldots,T,
\EEAS
with periods
\[
P_1 = 2100, \quad P_2 = 1500, \quad P_3 = 1100, \quad P_4 = 1700,
\]
and coefficients
\[
a = -.17, \quad b = .23, \quad c = -.12, \quad d = .13.
\]
(There is no special significance to the specific form; this is 
just a simple way to generate a smoothly varying sequence.)
Then we generate data as $x_t  = \exp z_t$, with 
$z_t\sim \mathcal{N}(\mu_t,\sigma_t^2)$.
The `true' quantiles are then
\[
\exp\left(\mu_t + \sigma_t \Phi^{-1}(\eta)\right)    
, \qquad t=1, \ldots, T,\quad \eta = .15,.5,.85,
\]
where $\Phi$ is the cumulative distribution function of a standard normal
random variable.

\paragraph{Parameters.} The half-life is $H=100$, and the buffer sizes are $M=100$
and $M^\text{tail}=3M$. We sample $L=10$ points for the tail approximation.

\begin{figure}
    \centering
    \includegraphics[width=1.0\textwidth]{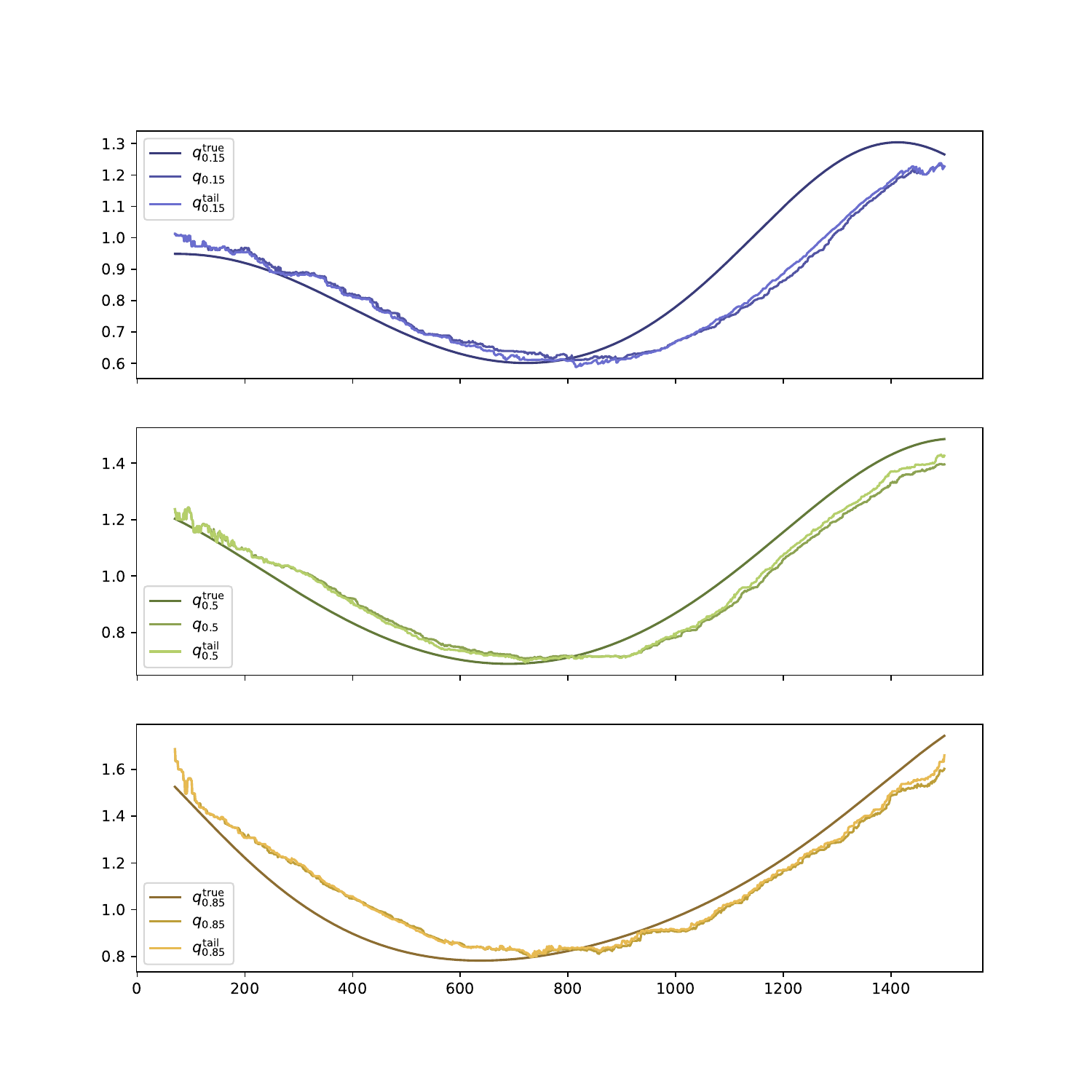}
    \caption{True and estimated quantile values across time using the exact and
    approximate finite memory EWMM methods.}
    \label{f-quantile}
\end{figure}

\begin{figure}
    \centering
    \includegraphics[width=1.0\textwidth]{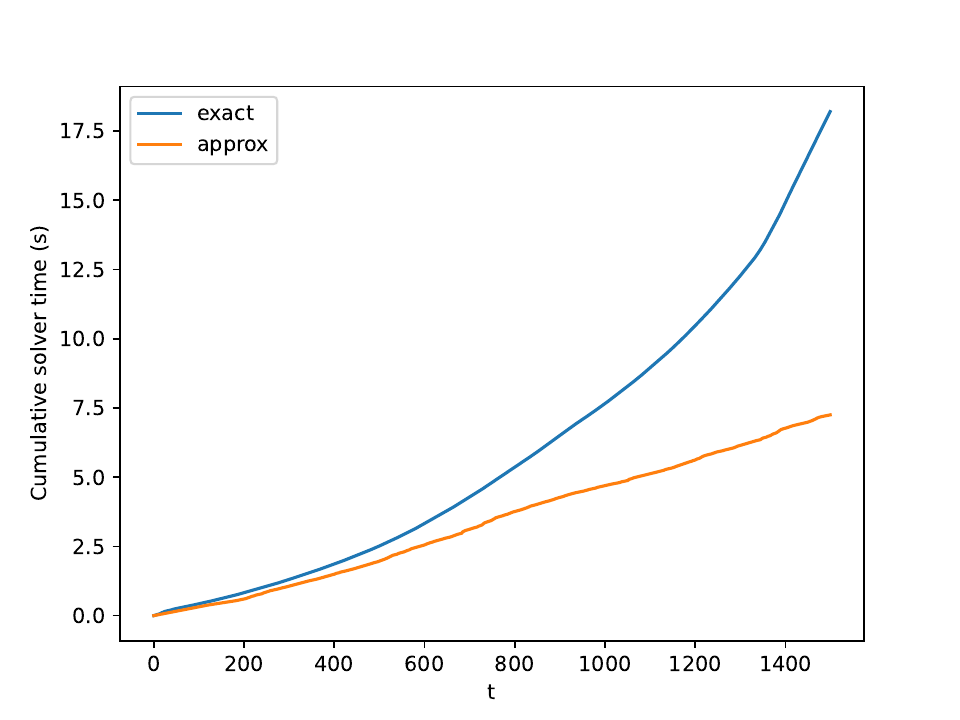}
    \caption{Time to fit the quantile estimation model using the exact and
    approximate finite memory EWMM methods.}
    \label{f-quantile-time}
\end{figure}

\begin{figure}
    \centering
    \includegraphics[width=1.0\textwidth]{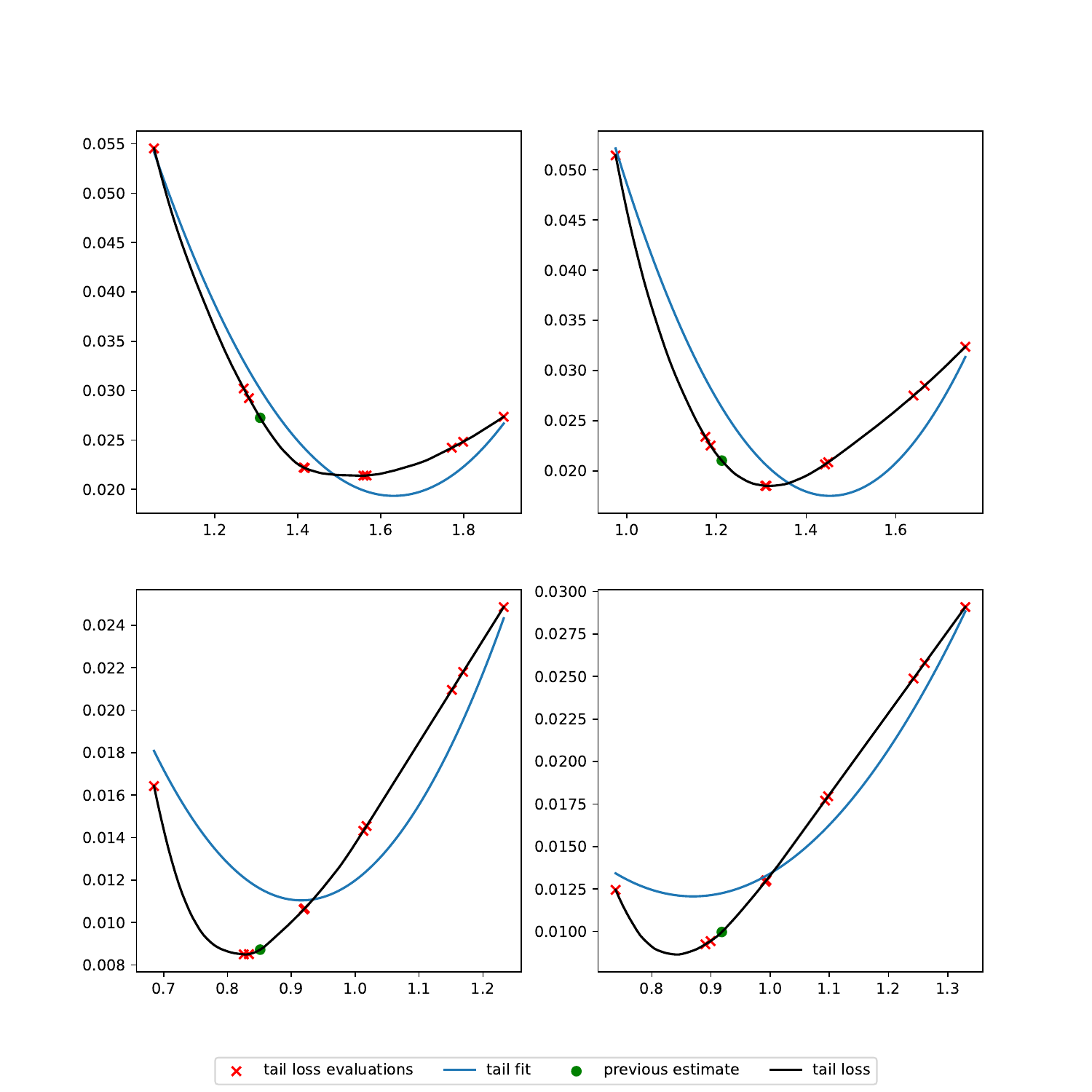}
    \caption{Quadratic tail approximations for the pinball loss.}
    \label{f-quantile-tail-fit}
\end{figure}

\paragraph{Results.}
We see that the approximate finite memory EWMM is able to closely match the results of
the exact method while incurring a fraction of the computational cost. We plot
the true quantile value and the estimated quantile values across time in
figure~\ref{f-quantile}. We also show how the computational effort of the two
methods compare in figure~\ref{f-quantile-time}. 
We show four examples of the quadratic tail approximations in figure~\ref{f-quantile-tail-fit}.


\clearpage
\subsection{Logistic regression}\label{s-logistic}
In this example, we generate data from a joint distribution
of features and targets, and use the approximate finite memory EWMM for a
logistic regression model to make predictions. We use the logistic loss function
\eqref{e-logistic-loss} and regularizer $r(\theta) = \lambda \|\theta\|_2^2$. 
We fit the approximate finite memory EWMM using the recursive Taylor
approximation method described in \S\ref{s-re-taylor}.

\paragraph{Data.}
We first generate a smoothly varying sequence of parameters
$\theta_t^\text{true}\in \reals^3$
as
\[
    \theta_t = a \sin(2\pi t/4000) +b\cos(2\pi t/6000), \quad t=1,\ldots,T=2000,
    \]
    where \[
        a = \begin{bmatrix}
            .17\\.23\\-.12
        \end{bmatrix},\quad
        b = \begin{bmatrix}
            .13\\-.11\\-.19
        \end{bmatrix}.
\]
We then generate pairs $x_t = (z_t,y_t)$ for $t=1,\ldots,T$ as independent samples from the following joint
distribution parametrized by $\theta_t^\text{true}$:
\[
    z_t \sim \mathcal{N}(\theta_t^\text{true}, I), \qquad y_t = \sign(z_t^T\theta_t + \xi_t), \qquad\xi_t \sim \mathcal{N}(0,\sigma^2).
\]
We use $\sigma=.1$. 

\paragraph{Parameters.}
We take $\lambda = 0.5$, half-life $H=150$, and $M=H$.

\begin{figure}
    \centering
    \includegraphics[width=1.0\textwidth]{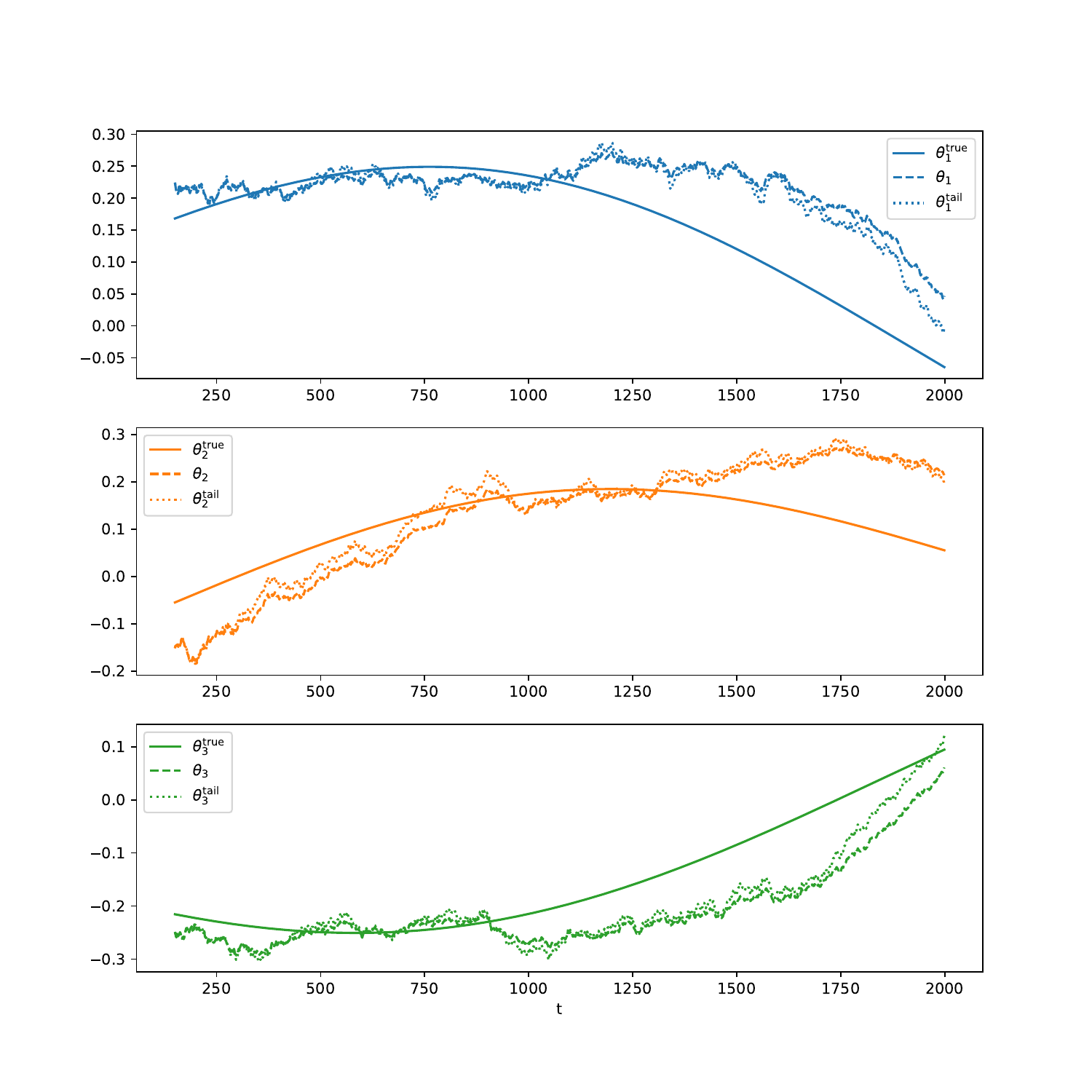}
    \caption{True and estimated parameters for the logistic regression model.}
    \label{fig:logistic-thetas}
\end{figure}

\begin{figure}
    \centering
    \includegraphics[width=1.0\textwidth]{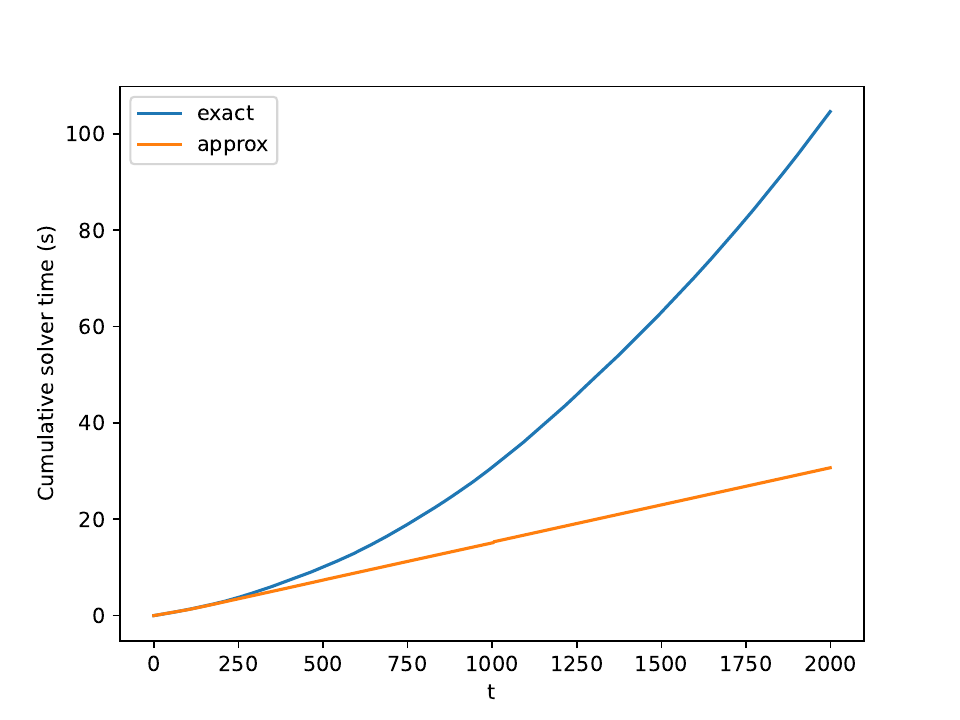}
    \caption{Time to fit the logistic regression model using the exact and
    approximate finite memory EWMM methods.}
    \label{fig:logistic-time}
\end{figure}

\paragraph{Results.} 
We show the true value $\theta_t^\text{true}$ and the estimate $\theta_t$
across time for the full and tail approximation models in figure~\ref{fig:logistic-thetas}.
We see that the approximate finite memory EWMM is
able to closely match the performance from the exact EWMM. We show the cumulative time to fit
the model across time in figure~\ref{fig:logistic-time}.

\clearpage
\section{Conclusions}
We have introduced the general exponentially weighted moving model,
which generalizes the well-known exponentially weighted moving average.
The idea is simple, and closely related to other well-known methods.

When the loss is quadratic, a simple recursion can be used to exactly compute
the EWMM estimate by forming and solving a fixed size problem.  When the 
parameters are from an exponential family, we compute the EWMA of
the sufficient statistic.  This special case includes some obvious ones,
such as least squares regression (possibly with nonquadratic regularizer),
and some less obvious ones like sparse inverse covariance.

When the loss is not quadratic, a simple recursion cannot be used. Instead
we propose an approximate method that stores a fixed window of data and carries
out computation that does not grow with time.

In this paper we do not suggest or recommend EWMMs for applications;
we simply address the question of how to compute it, 
or an approximation of it, efficiently.

\section*{Acknowledgements}
The authors thank Mykel Kochenderfer and Trevor Hastie for their helpful suggestions.

\clearpage
\printbibliography

\clearpage
\end{document}